\documentclass[11pt,draftcls,peerreview]{IEEEtran}

\usepackage{latexsym}
\usepackage{theorem}
\usepackage[dvips]{graphicx}

\title{Asymptotically Optimal Tree-based Group Key Management Schemes}

\author{\authorblockN{Hideyuki Sakai}\\
{\small
\authorblockA{Hitachi, Ltd., Systems Development Laboratory  \\
Asao-ku, Kawasaki-shi,
Kanagawa 215--0013, Japan\\
Email: sakai@sdl.hitachi.co.jp}}\\
\and\vspace{0.5cm}
\authorblockN{Hirosuke Yamamoto}\\
{\small 
\authorblockA{School of Frontier Schiences,
University of Tokyo\\
Kashiwa-shi, 
Chiba 277--8561, Japan\\
Email: Hirosuke@ieee.org}}
}

\theoremstyle{plain}
\theorembodyfont{\rmfamily}
\newtheorem{thm}{Theorem}

\theoremstyle{break}
\theorembodyfont{\rmfamily}
\newtheorem{algo}{Algorithm}

\theoremstyle{break}
\theorembodyfont{\rmfamily}

\theoremstyle{break}
\theorembodyfont{\rmfamily}

\theoremstyle{plain}
\theorembodyfont{\rmfamily}
\newtheorem{lem}{Lemma}

\begin{document}
\maketitle

\begin{center}{\Large ({\em Submitted to the IEEE Transactions on Information Theory})}
\end{center}
\vspace{0.5cm}

\begin{abstract}
In key management schemes that realize secure multicast communications 
encrypted by group keys on a public network,  tree structures are often used 
to update the group keys efficiently. 
Sel\c{c}uk and Sidhu have proposed an efficient scheme which updates dynamically the tree structures based on the withdrawal probabilities of members. 
In this paper, it is shown that
Sel\c{c}uk-Sidhu scheme is asymptotically optimal for the cost of withdrawal. 
Furthermore, a new key management scheme,
which takes account of key update costs of joining in addition to withdrawal,
is proposed. It is proved that the proposed scheme is also asymptotically optimal,
and it is shown by simulation that it can attain good performance for nonasymptotic cases.

\end{abstract}


\section{Introduction}

In the multicast communication of a group on a public network, 
a group secret key is often used to realize secure communication.
But, when a member joins and/or withdraws from the group, 
a new group key must be redistributed.

The Logical Key Hierarchy (LKH) scheme,
which was independently proposed by Wallner-Harder-Agee \cite{Wallner} 
and Wong-Gouda-Lam \cite{Wong} in 1997, is a scheme with a tree structure 
that can renew the group key securely and efficiently when a member changes.
Poovendran and Baras \cite{Poovendran} analyzed the LKH scheme information-theoretically 
by considering the withdrawal probability of members in the scheme.
Furthermore, Sel\c{c}uk and Sidhu \cite{Selcuk} have proposed a more efficient scheme 
such that  a tree structure is dynamically updated based on the withdrawal probabilities 
of members. They analyzed the performance of their scheme information-theoretically. 
But, their evaluation is very loose.

In this paper, we derive an asymptotically tight upper bound of the key update cost in 
Sel\c{c}uk-Sidhu scheme.
More precisely, the key update cost is $O(\log n)$ when a group has $n$ members, and
our upper bound is tight  within a constant factor which does not depend on $n$.
Furthermore, we propose a new dynamical key management scheme, 
which takes account of key update costs for joining in addition to withdrawal.
We show that the proposed scheme is also asymptotically optimal. 
Moreover, it is shown by simulation that  in nonasymptotic cases, the proposed scheme is more
efficient than Sel\c{c}uk-Sidhu scheme for joining 
while it is almost as efficient as Sel\c{c}uk-Sidhu scheme for
withdrawal.

In this paper, we assume that channels are noiseless and public.
Hence, any information sent over the channels may be wiretapped 
by adversaries who may be inside or outside of the group.
Each member has a private key and several subgroup keys 
in addition to a group key. The subgroup key and group key are shared 
by the members of a  subgroup and the group, respectively. 
The group key is used to encrypt secret messages to communicate 
among the group. On the other hand, the private key and subgroup keys 
are used when the keys must be updated by the change of members.

Furthermore, we suppose the following in this paper.
A reliable server, who has all the keys in the group, updates and 
distributes new keys when a member changes.
The number of members in the group is sufficiently large, 
and the frequency of joining and withdrawal is relatively large. 
The key update cost is evaluated by the number of keys that must be 
updated when a member changes.
To keep the security of communication, the key management scheme needs to 
meet the so-called Forward Security and Backward Security, which are
defined as follows.

\begin{itemize}
	\item{[Forward Security]}  A member who withdraws from a group 
	cannot decrypt any data that will be sent in the group after the withdrawal. 
	\item{[Backward Security]} A member who joins a group cannot decrypt 
	the data that were sent in the group before the joining.
\end{itemize}

In Section II,  Sel\c{c}uk-Sidhu scheme is reviewed, 
and the performance of the scheme is evaluated precisely in Section III.
Furthermore, in Section IV, Sel\c{c}uk-Sidhu scheme is extended to consider the cost of joining. 
Finally, some simulation results are shown in Section V.

\section{Sel\c{c}uk-Sidhu scheme}

The LKH scheme \cite{Wallner,Wong} can be represented by a binary tree
such that each member of a group corresponds to each leaf of the tree while
the root, each internal node, and each leaf also correspond to the group key,
a subgroup key, and a private key, respectively.
Each member holds all the keys on the path from the root to the leaf of the 
member in the tree.
Each internal node makes a subgroup which consists of the descendants of the 
node, and the subgroup can communicate securely against any other members
not included in the subgroup by using the subgroup key.
In the multicast communication of the group,  the group key is used
to realize secure communication. But, when a member joins or withdraws from 
the group, the subgroup keys and private keys are used
to update the keys.
Note that in order to keep security, it is necessary to update all the keys
on the path from the root to the leaf of the member.

For the LKH scheme, Poovendran and Baras \cite{Poovendran} 
introduced the withdrawal probabilities of members
to analyze information-theoretically the average cost of key update
in the case of the withdrawal.
Let ${\cal G}$ be a group and let
$P_M$ be the probability that a member $M\in{\cal G}$ withdraws from the group within a certain period\footnote{%
In order to keep the system securely, all keys are usually renewed periodically. Hence, the period is finite and
$P_M<1$ for many members.}. 
$P_M$ is assumed to be given since it can be often 
estimated from the statistics and the personal data of the member.
$P_M$ satisfies $0<P_M\le 1$. But, note that
\begin{equation}
P_{\cal G} \equiv \sum_{M\in{\cal G}} P_M, \label{defPr}
\end{equation}
is usually not equal to one. Hence, we use the normalized withdrawal
probability distribution  ${\cal P}\equiv \{P_M/P_{\cal G}: M\in {\cal G}\}$ to evaluate the performance.

When a member withdraws from the group,
the average withdrawal cost $L$ and the average normalized withdrawal cost 
$l$ are defined by
\begin{eqnarray}
L & \equiv& \sum_{M\in{\cal G}} P_M d_M, \label{pdn-1}\\
l & \equiv& \sum_{M\in{\cal G}}\frac{P_M}{P_{\cal G}} d_M, \label{pdn} 
\end{eqnarray}
respectively, 
where $d_M$ is the number of keys that must be updated when member $M$ withdraws. We note that $d_M$ is equal to the
depth of member $M$ in the key tree of the LKH scheme.

In the case of lossless source coding, 
$l$ given by (\ref{pdn}) corresponds to the average code length 
for a fixed-to-variable length code (FV code) with 
probability distribution ${\cal P}$ and codeword length $\{d_M: M\in{\cal G}\}$, and it is well known that the Huffman tree \cite{Huffman} is 
the best tree to minimize the average code length under the prefix condition.
Furthermore, if the group is incremented and the probability distribution changes 
as the coding progresses, the optimal code tree can be kept by 
the dynamic Huffman coding algorithm \cite{Faller}\cite{Gallager}.

In the case of key management, the prefix condition is also required because
the set of keys of each member must be different from that of others
to keep security. 
Based on this observation, 
Poovendran and Baras have shown that in the case of key management, 
the Huffman tree is the best tree to minimize the average normalized withdrawal cost.
However,  if the key tree is updated by the dynamic Huffman coding algorithm
to keep the key tree optimally, the key update cost cannot be minimized 
usually because the algorithm often changes the tree structure for many members
besides a withdrawn member, and this causes additional key update costs.
Hence, in the case of key management, it is better to keep the tree structure
as unchanged as possible for non-withdrawn members.
Based on this idea,  Sel\c{c}uk and Sidhu \cite{Selcuk} have proposed 
two key tree updating algorithms.

\begin{figure}[t]
\begin{center} 
\includegraphics[width=.78\linewidth]{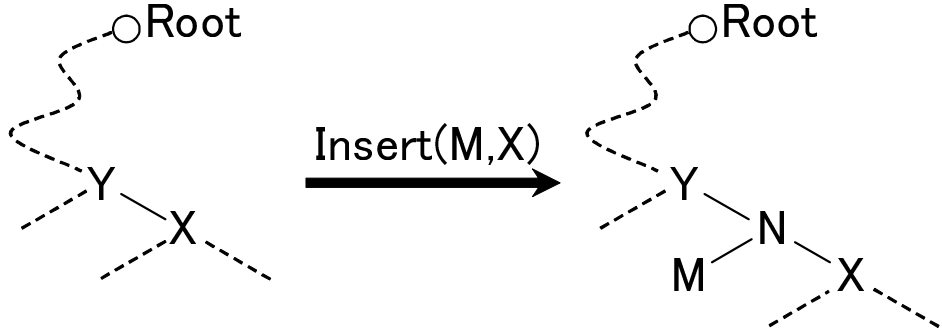} 
\caption{Insertion of $M$ by Insert$(M,X)$.}\label{fig-1} 
\end{center}
\end{figure}

In order to explain Sel\c{c}uk-Sidhu algorithms, 
we first define an operation Insert$(M,X)$,
which represents the insertion of a new member $M$ at node $X$,
i.e.~a new node $N$ is inserted between $X$ and its parent node $Y$ 
as shown in Fig.~\ref{fig-1}, and $M$ is linked as a child of $N$. 

For node $X$, let $P_X$ be the weight that is given by 
the sum of the withdrawal probabilities of all members included 
in the descendants of node $X$.\footnote{If $X$ is the root, 
$P_X$ is equal to $P_{\cal G}$.}
Then, the first algorithm to update a key tree is described as follows. 

\begin{algo}\label{Join1}

Let $M$ be a new member and let $X$ be the root of a given key tree.
\begin{enumerate}
	\item[1.] If $X$ is a leaf, then operate Insert$(M,X)$ and exit.
	\item[2.] Let $X_l$ and $X_r$ be the left and right children of $X$,
	         respectively. 
	If it holds that $P_M \ge P_{X_l}$  and $P_M \ge P_{X_r}$, then 
	operate Insert$(M,X)$ and exit.
	\item[3.] If $P_{X_l} \ge P_{X_r}$, then let $X\leftarrow X_r$.  Otherwise,
	let $X\leftarrow X_l$. Go back to Step 1.
\end{enumerate}
\end{algo}

In order to describe the second algorithm, we first define 
the cost increase $C_{M,X}$ for a new member $M$ and
a node $X$ as follows \cite{Selcuk}.

\begin{equation}
C_{M,X} \equiv (d_X+1)P_M + P_X, \label{cost} 
\end{equation}
which represents the increase of cost $L$
for the case that a new member $M$ is inserted at node $X$.

Let $C_{\min}$ be the minimum cost increase that is given by
\begin{equation}
C_{\min} \equiv \min_X C_{M,X}. 
\end{equation}
Then, the second algorithm inserts a new member $M$ at the
node that can attain $C_{\min}$. Formally, the second algorithm to 
update a tree key is defined as follows.

\begin{algo}\label{Join2}
Let $M$ be a new member .
\begin{enumerate}
  \item First calculate $C_{M,X}$ for every node $X$, and obtain $C_{\min}$. 
           Let $X_{\min}$ be the node that attains $C_{\min}$.
  \item Operate Insert$(M, X_{\min})$.
\end{enumerate}
\end{algo}

It is shown by simulation in  \cite{Selcuk} 
that Algorithm 2 can attain less average withdrawal cost $L$ than 
Algorithm 1. But, although Algorithm 1 can be implemented with $O(\log n)$ time complexity 
when $|{\cal G}|=n$, 
i.e.~the size of a group is $n$, Algorithm 2 requires $O(n)$ time
complexity in the search of $X_{\min}$.

Sel\c{c}uk and Sidhu evaluated the average normalized withdrawal cost $l$ for 
the case of Algorithm 1 as follows\footnote{In this paper, the base of $\log$ is $2$.} \cite{Selcuk}.
\begin{eqnarray}
d_M &\le& K_1(- \log P_M + \log P_{\cal G} ) +K_2, \label{dxselcuk} \\
l &\le& K_1H({\cal P}) +K_2, \label{oldpd}
\end{eqnarray}
where $H({\cal P})$ is the entropy of the probability distribution ${\cal P}=\{P_M/P_{\cal G}: M\in{\cal G}\}$,
and it is defined by
\begin{equation}
H({\cal P}) \equiv 
   -\sum_{M\in{\cal G}} \frac{P_M}{P_{\cal G}}\log \frac{P_M}{P_{\cal G}}. \label{entropy}
\end{equation}
$K_1$ and $K_2$ are constants given by
\begin{eqnarray}
K_1 &\equiv& \frac{1}{\log \alpha} \approx 1.44, \label{K1}\\
K_2 &\equiv& \frac{1}{\log \alpha}\log\frac{\sqrt{5}}{\alpha} \approx 0.672,
\end{eqnarray}
where $\alpha = \frac{1+\sqrt{5}}{2}$.

We note from the source coding theorem for FV codes \cite{Cover} that
the average normalized withdrawal cost $l$ must satisfy
\begin{equation}
 l \ge H({\cal P}).  \label{lower}
\end{equation}
Furthermore,  it holds from Theorem \ref{Hzouka} shown below
that $H({\cal P})= O(\log n)$ for $|{\cal G}|=n$.
Hence, 
the upper bound of $l$ given by (\ref{oldpd}) is not asymptotically tight
as $n$ becomes large.
This result means that Algorithm 1 is not efficient or the upper bound is loose. 
In the next section, we will show that Algorithm 1 is asymptotically optimal
by deriving an asymptotically tight upper bound.

\begin{thm}\label{Hzouka}
Assume that the maximum and minimum probabilities of withdrawal defined by
\begin{eqnarray}
P_{\max} &\equiv& \max_{M\in{\cal G}}P_M \le 1, \label{Pmax}\\
P_{\min} &\equiv& \min_{M\in{\cal G}}P_M >0
\end{eqnarray}
are fixed. Then, for $n=|{\cal G}|$,  $H({\cal P})$ defined by (\ref{entropy}) satisfies
\begin{eqnarray}
H({\cal P}) = O(\log n). \label{Hinf} 
\end{eqnarray}
\end{thm}

\begin{proof} 
Let
$\epsilon_{\min} = P_{\min}/P_{\cal G}$, $\epsilon_{\max} = P_{\max}/P_{\cal G}$, 
and $k=\epsilon_{\min}/\epsilon_{\max}=P_{\min}/P_{\max}$.
Then, $H({\cal P})$ can be bounded as follows.
\begin{eqnarray}
H({\cal P}) 
 &=& -\sum_{M\in{\cal G}}\frac{P_M}{P_{\cal G}}\log\frac{P_M}{P_{\cal G}} \nonumber \\
&\stackrel{(a)}{\ge}& -\sum_{M\in{\cal G}}\epsilon_{\min}\log\epsilon_{\min} \nonumber \\
&=& n\epsilon_{\min}\log\frac{1}{\epsilon_{\min}} \nonumber \\
&\stackrel{(b)}{\ge}& \frac{\epsilon_{\min}}{\epsilon_{\max}}\log n \nonumber \\
&=& k\log n \label{klog},
\end{eqnarray}
where inequalities $(a)$ and $(b)$ hold because of the following reasons. 
\begin{description}
\item[$(a)$:]$-t\log t$ is monotonically increasing when $t>0$ is small. Furthermore,
when $n$ is sufficiently large and $P_{\cal G} \gg 1$, we have that 
$\epsilon_{\min}=\frac{P_{\min}}{P_{\cal G}}\leq \frac{1}{P_{\cal G}} \ll 1$.
\item[$(b)$:]From the relation $nP_{\max} \ge P_{\cal G}  \ge nP_{\min}$,
it holds that
\begin{equation}
\frac{1}{\epsilon_{\min}}= \frac{P_{\cal G}}{P_{\min}} \ge n 
   \ge \frac{P_{\cal G}}{P_{\max}}=\frac{1}{\epsilon_{\max}}.
\end{equation}
\end{description}

Similarly, we can easily show that
\begin{eqnarray}
H({\cal P})  \le \frac{1}{k}\log n.
\label{klog-2}
\end{eqnarray}
Therefore, (\ref{Hinf}) is obtained from (\ref{klog}) and (\ref{klog-2}).
\end{proof}

\section{Detailed Analysis of Sel\c{c}uk-Sidhu scheme}

In order to derive a tight upper bound for the key tree constructed 
by Algorithm 1,  we use the following lemma.

\begin{lem}\label{sib}
Let  $X$  and $S$ be sibling nodes each other in the key tree constructed 
by Algorithm \ref{Join1}. Then, it holds that
\begin{eqnarray}
|P_X - P_S| \le P_{\max}, \label{the1}
\end{eqnarray}
where $P_{\max}$ is defined in (\ref{Pmax}).
\end{lem}

\begin{proof}
The lemma can be proved by mathematical induction for the key tree with $|{\cal G}|=n$.
Let $P_X^{(n)}$, $P_S^{(n)}$, and $P_{\max}^{(n)}$ be $P_X$, $P_S$, and $P_{\max}$ in
the case of $|{\cal G}|=n$, respectively.
\begin{itemize}
	\item[1.] When $n=2$, it holds that $P_X^{(2)} \le P_S^{(2)} = P_{\max}^{(2)}$ or 
	$P_S^{(2)} < P_X^{(2)} = P_{\max}^{(2)}$. In the former case, we have 
	$0\leq P_S^{(2)} - P_X ^{(2)}< P_{\max}^{(2)}$.
	Otherwise, $0<P_X^{(2)} - P_S ^{(2)}< P_{\max}^{(2)}$. Hence,  (\ref{the1}) holds.
	\item[2.] Supposed that 
	\begin{equation}
	|P_X^{(n)} - P_S^{(n)}| \le P_{\max}^{(n)}  \label{eq-L1-1}
	\end{equation}
	holds for every pair of sibling nodes $(X, S)$ in the key tree with $|{\cal G}|=n$, 
	and the key tree is incremented to $|{\cal G}|=n+1$ by inserting a new member $M$ 
	with probability $P_M$ according to Algorithm \ref{Join1}. Then, we have
	\begin{equation}
	P_{\max}^{(n+1)}=\max\{P_{\max}^{(n)}, P_M\}\geq P_{\max}^{(n)}. \label{eq-L1-2}
   \end{equation}	
   We assume, without loss of generality,  that $P_S^{(n)}  \ge P_X^{(n)} $. Then, 
   from Algorithm  \ref{Join1}, 
   there may occur the following three cases.
\begin{itemize}
  \item[Case 1: ]
  
  $M$ is inserted outside nodes $X$, $S$, and their descendants. 
    
   In this case, it  holds obviously that $P_X^{(n+1)}=P_X^{(n)}$
   and $P_S^{(n+1)}=P_S^{(n)}$.
	Hence, we obtain from (\ref{eq-L1-1}) and (\ref{eq-L1-2}) that 
	$|P_X^{(n+1)} - P_S^{(n+1)}| \le P_{\max}^{(n+1)}$.
 \item[Case 2:]
 
  $M$ is inserted at node $X$ as shown in Fig.~\ref{fig-1}. 
	
	In this case, we have the new pairs of sibling nodes, $(X, M)$ and $(N, S)$, 
	where $N$ was the new parent node of $X$, and it holds from
	Step 2 of Algorithm \ref{Join1} that [$P_X^{(n)}\leq P_M<P_S^{(n)}$ or
	$P_M< P_X^{(n)}\leq P_S^{(n)}$] and [$P_M \geq P_{X_l}^{(n)}$,  
	$P_M \geq P_{X_r}^{(n)}$], where $X_l$ and $X_r$ are the children of
	$X$. Hence, from $P_X^{(n+1)}=P_X^{(n)}$  and $P_S^{(n+1)}=P_S^{(n)}$,
	we have that [$P_X^{(n+1)}\leq P_M<P_S^{(n+1)}$ or
	$P_M< P_X^{(n+1)}\leq P_S^{(n+1)}$] and $P_X^{(n+1)}\leq 2P_M$. 
	
	In the case of $P_X^{(n+1)}\leq P_M$, it holds that 
	$0\leq P_M-P_X^{(n+1)} < P_S^{(n+1)}-P_X^{(n+1)}=P_S^{(n)}-P_X^{(n)}\leq P_{\max}^{(n)}\leq P_{\max}^{(n+1)}$.
	Furthermore, in the case of $P_X^{(n+1)}>P_M$, 
	it holds that $0<P_X^{(n+1)}-P_M \leq 2P_M-P_M=P_M\leq P_{\max}^{(n+1)}$. 
	
	For the pair $(N,S)$, we have that $|P_S^{(n+1)}-P_N^{(n+1)}| =|P_S^{(n+1)}-P_X^{(n+1)}-P_M|=
	|(P_S^{(n)}-P_X^{(n)})-P_M| \leq \max\{P_{\max}^{(n)}, P_M\}=P_{\max}^{(n+1)}$.
\item[Case 3:]

 $M$ is inserted at a descendant node of $X$.
         
         In this case, we have that $P_S^{(n+1)}=P_S^{(n)}$ and $P_X^{(n+1)}=P_X^{(n)}+P_M$
	Hence,  it holds that 
	$|P_S^{(n+1)}-P_X^{(n+1)}| =|(P_S^{(n)}-P_X^{(n)})-P_M| \leq \max\{P_{\max}^{(n)}, P_M\}=P_{\max}^{(n+1)}$.
	
\end{itemize}

\end{itemize}
\end{proof}

Now, we evaluate the weight of the ancestors of an arbitrarily given node $X$ 
in the key tree generated by Algorithm 1.
 Let nodes $F$ and $G$ be the parent and grandparent 
of $X$, respectively, and let $U$ be the sibling of $F$.
Then, we have from Lemma \ref{sib} that 
\begin{eqnarray}
P_F &=& P_X + P_S \nonumber \\
   &\ge& 2P_X - P_{\max}. \label{ys} 
\end{eqnarray}
Furthermore, we have that
\begin{eqnarray}   
P_G &=& P_F + P_U \nonumber \\
      &\ge& 2P_F - P_{\max} \nonumber \\
      &\ge& 2^2P_X - 2P_{\max} - P_{\max}, \label{fu}
\end{eqnarray}
where the first and second inequalities holds from
Lemma~\ref{sib} and  (\ref{ys}), respectively.
By repeating the same procedure, we obtain that
\begin{eqnarray}
P_{\cal G} &\ge& 2^{d_X}P_X - (2^{d_X-1} + 2^{d_X-2} + \cdots + 2 + 1)P_{\max} \nonumber \\
&=& 2^{d_X}(P_X - P_{\max}) + P_{\max}, \label{Prr}
\end{eqnarray}
where $d_X$ is the depth of node $X$.

Therefore, the following theorem holds.

\begin{thm}
In the key tree constructed by Algorithm 1, 
the following relation holds for any node $X$ and any leaf $M_X$ that is a descendant of  $X$.
\begin{eqnarray}
d_X & \le & \log P_{\cal G} + \log \left(1 - \frac{P_{\max}}{P_{\cal G}} \right) 
             - \log(P_X - P_{\max}) \label{l1} \\
d_{M}^{(X)} &\le& K_1( - \log P_{M} + \log P_X )+ K_2, \label{l2}
\end{eqnarray}
where $d_M^{(X)}$ is the depth from node $X$ to leaf $M$.
\end{thm}
\begin{proof}
(\ref{l1}) and (\ref{l2}) hold from (\ref{Prr}) and (\ref{dxselcuk}), respectively.
\end{proof}

Next, we evaluate $P_X$.
\begin{lem}\label{py}
Let $X_l$ and $X_r$ be the children of node $X$.
Assume that the weight of $X$ is larger than a real number $t$ but
the weight of $X_l$ is not larger than $t$, i.e.~$P_X>t\ge P_{X_l}$.
Then, the following inequalities hold.
\begin{eqnarray}
t < P_X \le 2t + P_{\max} \label{yhani}
\end{eqnarray}
\end{lem}
\begin{proof}
From (\ref{the1}), we obtain that
\begin{equation}
P_X = P_{X_l} + P_{X_r}  \le 2 P_{X_l} +P_{\max} \le 2t + P_{\max}. \label{PX}
\end{equation}
\end{proof}

Let $t (>P_{\max})$ be a parameter which will be optimized later.
Now, for a given leaf $M$, we consider the node $X$ that is the
nearest ancestor of $M$ under the condition $P_X>t$.
Then, from (\ref{l1}), (\ref{l2}), and (\ref{PX}), 
the depth $d_M$ of leaf $M$ can be bounded as follows.
\begin{eqnarray}
d_M &=& d_X + d_{M}^{(X)} \nonumber \\
&\leq& \log P_{\cal G} + \log \left( 1 - \frac{P_{\max}}{P_{\cal G}} \right)
 - \log(P_X - P_{\max}) 
 + K_1(- \log P_M + \log P_X) + K_2 \nonumber\\
&<& \log P_{\cal G} + \log \left( 1 - \frac{P_{\max}}{P_{\cal G}} \right)
 - \log(t - P_{\max}) \nonumber \\
&\phantom{=}& \hspace{5cm}+ K_1(- \log P_M + \log(2t + P_{\max})) + K_2 \label{dxtemp}
\end{eqnarray}

We can easily show for $f(t) = - \log(t - P_{\max}) + K_1 \log(2t + P_{\max})$
that $f(t)$ can be minimized at $t=t_m$ given by
\begin{eqnarray}
t_m= \frac{2 + \log \alpha}{2(1 - \log \alpha)} \approx 4.405, \label{argt}
\end{eqnarray}
where $\alpha = \frac{1+\sqrt{5}}{2}$.
Note that if a key tree is sufficient large and efficiently constructed, there exists the node $X$ that satisfies $P_X>t_m>P_{\max}$. Hence, by substituting $t=t_m$ into (\ref{dxtemp}) and some 
calculations, 
we obtain the following theorem.

\begin{thm}\label{dxup}
When the key tree constructed by Algorithm 1 is sufficiently large,
the depth $d_M$ of a leaf $M$ in the key tree is upper bounded as follows.
\begin{eqnarray}
d_M &<& \log P_{\cal G} - K_1\log P_M + (K_1 - 1)\log P_{\max} 
 + \log \left( 1 - \frac{P_{\max}}{P_{\cal G}} \right) + K_3, \label{dxnew}
\end{eqnarray}
where $K_3$ is defined by
\begin{eqnarray}
K_3 &= & - \log \frac{3\log\alpha}{2(1 - \log\alpha)} 
+\frac{1}{\log\alpha}\left( \log \frac{3}{1 - \log \alpha} 
+ \log \frac{\sqrt{5}}{\alpha}\!\right) \nonumber \\
&\approx& 3.65.
\end{eqnarray}
\end{thm}

By averaging $d_M$ for all member in ${\cal G}$, the following theorem 
holds for the average normalized withdrawal cost $l$. 

\begin{thm}\label{newLt}
When a key tree constructed by Algorithm 1 is sufficiently large,
the average normalized withdrawal cost $l$ of the key tree
satisfies that 
\begin{eqnarray}
l &<&  H({\cal P}) + (K_1 - 1)\log \frac{P_{\max}}{P_{\min}} 
+ \log \left( 1 - \frac{P_{\max}}{P_{\cal G}} \right) + K_3. \label{newL}
\end{eqnarray}
\end{thm}

\begin{proof}
$l$ can be evaluated as follows.
\begin{eqnarray}
l &=& \sum_{M\in{\cal G}} \frac{P_M}{P_{\cal G}} d_M \nonumber \\
  &<& \log P_{\cal G} - K_1\sum_{M\in{\cal G}} \frac{P_M}{P_{\cal G}}\log P_M
   + (K_1 - 1)\log P_{\max} 
        + \log \left( 1 - \frac{P_{\max}}{P_{\cal G}} \right) + K_3 \nonumber \\
&=& \log P_{\cal G} - K_1\sum_{M\in{\cal G}}\frac{P_M}{P_{\cal G}}\log \frac{P_M}{P_{\cal G}} 
    - K_1\sum_{M\in{\cal G}}\frac{P_M}{P_{\cal G}}\log P_{\cal G} \nonumber \\
&\phantom{=} & \hspace{4.5cm}+ (K_1 - 1)\log P_{\max} 
     + \log \left( 1 - \frac{P_{\max}}{P_{\cal G}} \right) + K_3 \nonumber \\
&\stackrel{(a)}{=} & \log P_{\cal G} + K_1H({\cal P}) - K_1\log P_{\cal G} 
 + (K_1 - 1)\log P_{\max} 
    + \log \left( 1 - \frac{P_{\max}}{P_{\cal G}} \right) + K_3 \nonumber \\
&=& K_1H({\cal P}) - (K_1 - 1)\log P_{\cal G} 
 + (K_1 - 1)\log P_{\max} 
      + \log \left( 1 - \frac{P_{\max}}{P_{\cal G}} \right) + K_3 \nonumber \\
&\stackrel{(b)}{\le}& K_1H({\cal P}) 
+ (K_1 - 1)\left( \log \frac{1}{P_{\min}} - H({\cal P}) \right)  + (K_1 - 1)\log P_{\max} 
    + \log \left( 1 - \frac{P_{\max}}{P_{\cal G}} \right) + K_3 \nonumber \\
&=& H({\cal P}) + (K_1 - 1)\log \frac{P_{\max}}{P_{\min}} 
 + \log \left( 1 - \frac{P_{\max}}{P_{\cal G}} \right) + K_3, \label{newpd}
\end{eqnarray}
where equality $(a)$ and inequality $(b)$ hold from (\ref{entropy}) and the following lemma,
respectively.
\end{proof}

\begin{lem}\label{hpr}
$H({\cal P})$, $P_{\cal G}$, and $P_{\min}$ satisfy that
\begin{eqnarray}
- \log P_{\cal G} &\le& \log \frac{1}{P_{\min}} - H({\cal P}). \label{ineq} 
\end{eqnarray}
\end{lem}
\begin{proof}
It is well known that the entropy $H({\cal P})$ is bounded by $\log n$
for $|{\cal G}|=n$, and it holds obviously that $P_{\cal G} \ge nP_{\min}$.
Hence, we obtain that
\begin{eqnarray}
H({\cal P}) - \log P_{\cal G} \le \log n - \log \left( nP_{\min} \right) = \log \frac{1}{P_{\min}}.
\end{eqnarray}
\end{proof}

We finaly note that in (\ref{newL}), the coefficient of 
$H({\cal P}) =O(\log n)$ is one and the second and third terms are constants.
Hence, Theorem \ref{newLt} gives an asymptotically tight bound of $l$.

\section{Extension of Sel\c{c}uk-Sidhu scheme}

In Sel\c{c}uk-Sidhu scheme \cite{Selcuk}, only the withdrawal cost of a new member 
is considered. But, the withdrawal cost is an expected cost in the future, which may
not be occur. On the other hand, it is always necessary to update a key tree 
when a new member joins.
Hence, in this section, we propose extended schemes of Algorithms 1 and 2 to
consider the joining cost in addition to the withdrawal cost.

When a new member is inserted at the node $X$ with depth $d_X$, the 
withdrawal cost $L$ increases by $C_{M,X}$, which is given by (\ref{cost}).
But,  at the same time, $d_X + 1$ keys in the tree must be updated 
with probability one for the joining.
Hence,  the cost increase including the joining cost, say $C_{M,X}^*$,  
can be given by 
\begin{eqnarray}
C_{M,X}^* &\equiv& (d_X + 1) P_M + P_X + 1 \cdot (d_X + 1) \\
&=& (d_X + 1) (P_M + 1) + P_X. \label{newcost} 
\end{eqnarray}
 
 Comparing $C_{M,X}^*$ with $C_{M,X}$, 
we note that $P_M$ in $C_{M,X}$ is changed to $P_M+1$ in $C_{M,X}^*$.
Hence, by substituting $P_M+1$ into $P_M$ in Algorithms 1 and 2,
we can obtain the following algorithms which consider the joining cost.

\begin{algo}\label{Join3}

Let $M$ be a new member and let $X$ be the root of a given key tree.
\begin{enumerate}
	\item If $X$ is a leaf, then operate Insert$(M,X)$ and exit.
	\item Let $X_l$ and $X_r$ be the left and right children of $X$,
	         respectively. 
	If it holds that $P_M+1 \ge P_{X_l}$  and $P_M+1 \ge P_{X_r}$, then 
	operate Insert$(M,X)$ and exit.
	\item If $P_{X_l} \ge P_{X_r}$, then let $X\leftarrow X_r$.  Otherwise,
	let $X\leftarrow X_l$. Go back to Step 1.
\end{enumerate}
\end{algo}

\begin{algo}\label{Join4}
Let $M$ be a new member .
\begin{enumerate}
  \item First calculate $C_{M,X}^*$ for every node $X$, and obtain $C_{\min}^*$,
          where $\displaystyle C_{\min}^* \equiv \min_X C_{M,X}^*$. 
           Let $X_{\min}^*$ be the node that attains $C_{\min}^*$.
  \item Operate Insert$(M, X_{\min}^*)$.
\end{enumerate}
\end{algo}

For Algorithm~\ref{Join3}, the following theorem
holds in the same way as Theorem~\ref{newLt}.

\begin{thm}\label{newLkt}
When the key tree constructed by Algorithm 3 is sufficiently large,
the average normalized withdrawal cost $l$ of the key tree
satisfies that 
\begin{eqnarray}
l &<& H({\cal P}) + \log P_{\max} + (K_1 -1)\log(3P_{\max}+5) \nonumber \\
   &\phantom{=}& \hspace{2cm}- K_1\log P_{\min}+ \frac{P_{\max}+4}{P_{\min}}
     + \log \left( 1 -\frac{P_{\max}+2}{P_{\cal G}} \right) +K_4, \label{newLk} 
\end{eqnarray}
where $K_4$ is defined as follows.
\begin{eqnarray}
K_4 &=& - \left( \frac{1}{\log\alpha}-1 \right)\log\left(\frac{1}{\log\alpha}-1 \right) 
+\frac{1}{\log \alpha}\log \frac{2\sqrt{5}e}{\alpha\log e} \label{eq-th5-1} \\
&\approx& 3.95
\end{eqnarray}
\end{thm}
\begin{proof} (The proof is given in the appendix.)\end{proof}

We note from Theorem \ref{newLkt} that
the coefficient of $H({\cal P})$ in (\ref{newLk}) is also one although the constant terms are
larger than (\ref{newL}). This means that Algorithm 3 can
also attain asymptotically optimal key tree for the withdrawal cost in addition to decreasing
the joining cost.


\section{Simulation Results}

In the previous sections, we showed that Algorithms 1 and 3 are asymptotically optimal and Algorithms 2 and 4 are expected to 
achieve more efficient performance than Algorithms 1 and 3, 
respectively, in the case of withdrawal.  In this section, we evaluate
the performances of Algorithms 1--4 by simulation.

We first construct the optimal tree, i.e., Huffman tree for a group
with $n$ members.
Then, a new member joins the group each after a member withdraws
from the group. Such joining and withdrawal are repeated $m$ times.
It is assumed that the withdrawal probability of a new member $P_M$ is
uniformly distributed in $[0.1, 0.9]$.
For this case, 
the average costs of joining and withdrawal are shown in Tables~\ref{Simu1} and \ref{Simu2},
respectively.

\begin{table}[tb]
\caption{Average cases for Joining}\label{Simu1}
\begin{center}
\begin{tabular}{|c||c|c|c|c|}
\hline
$n$ & \multicolumn{2} {c|}{$100$} & \multicolumn{2} {c|}{$10,000$}\\
\hline
$m$ & $100$ & $10,000$ & $100$ & $10,000$ \\
\hline\hline
Alg.~\ref{Join1} & 7.42 & 7.32 & 13.19 & 14.14\\
\hline
Alg.~\ref{Join2} & 7.53 & 7.35 & 14.23 & 14.20\\
\hline
Alg.~\ref{Join3} & 6.50 & 6.23 & 12.85 & 13.12\\
\hline
Alg.~\ref{Join4} & 6.51 & 6.26 & 13.02 & 13.14\\
\hline
 \end{tabular}
\end{center}
\end{table}

\begin{table}[tb]
\caption{Average cases for Withdrawal}\label{Simu2}
\begin{center}
\begin{tabular}{|c||c|c|c|c|}
\hline
$n$ & \multicolumn{2} {c|}{$100$} & \multicolumn{2} {c|}{$10,000$}\\
\hline
$m$ & $100$ & $10,000$ & $100$ & $10,000$ \\
\hline\hline
Alg.~\ref{Join1} & 5.46 & 5.51 & 12.11 & 12.19\\
\hline
Alg.~\ref{Join2} & 5.39 & 5.45 & 12.03 & 12.14\\
\hline
Alg.~\ref{Join3} & 5.75 & 5.76 & 12.26 & 12.33\\
\hline
Alg.~\ref{Join4} & 5.57 & 5.71 & 12.13 & 12.28\\
\hline
 \end{tabular}
\end{center}
\end{table}

We note from the tables that
Algorithms \ref{Join3} and \ref{Join4} can improve the
cost of joining at a little increased cost of withdrawal.
Algorithms \ref{Join2} and \ref{Join4} are more 
efficient than Algorithms \ref{Join1} and \ref{Join3},
respectively, in the case of withdrawal. But the 
difference is not large, and Algorithms \ref{Join2} and \ref{Join4} 
require $O(n)$ time complexity although Algorithms \ref{Join1} and \ref{Join3} 
can be implemented with $O(\log n)$ time complexity.
Therefore, Algorithms \ref{Join3} and \ref{Join4} should be used in the cases
of large $n$ and small $n$, respectively. 

If the backward security described in section I is not required for a group, 
we don't need change any group and subgroup keys when a new member joins
the group.  Hence, it is preferable to use 
Algorithms \ref{Join1} or \ref{Join2} in such a case.

\appendix
\subsection{The proof of Theorem \ref{newLkt}}

For the key tree constructed by Algorithm \ref{Join3}, the following lemma holds.
\begin{lem}\label{Lemma-A1}
Let  $X$  and $S$ be sibling nodes each other in the key tree constructed 
by Algorithm \ref{Join3}. Then, it holds that
\begin{eqnarray}
|P_X - P_S| \le P_{\max}+2, \label{eq-A-1}
\end{eqnarray}
where $P_{\max}$ is defined in (\ref{Pmax}).
\end{lem}
\begin{proof} The lemma can be proved 
in the same way as Lemma \ref{sib}. \end{proof}

Now, for a give leaf $M$, let nodes $X$ and $Y$ be ancestors of $M$ such that $Y$ is an ancestor of $X$,
$P_X>1$, and $P_Y> P_{\max}+2$. When $|{\cal G}|=n$ is sufficiently large, there always exist such nodes $X$ and $Y$.
We represent the depths from the root to node $Y$, from node $Y$ to node $X$, and from node $X$  to leaf $M$ 
by $d_Y$, $d_X^{(Y)}$, $d_M^{(X)}$, respectively, which satisfy that 
\begin{equation}
d_M=d_Y+d_X^{(Y)}+ d_M^{(X)}.   \label{eq-A1-2}
\end{equation}
Then, by using Lemma \ref{Lemma-A1}, 
we can prove in the same way as  (\ref{l1}) and (\ref{l2}) that 
\begin{eqnarray}
  d_Y & \leq &\log P_{\cal G} + \log \left(1 - \frac{P_{\max}+2}{P_{\cal G}} \right) 
    - \log(P_Y - P_{\max}-2),  \label{eq-A-2}\\
 d_{X}^{(Y)} &\le& K_1[ - \log (P_{X}-1) + \log (P_Y -1) ]+ K_2. \label{eq-A3}
\end{eqnarray}
Furthermore,  $d_M^{(X)}$ obviously satisfies that
\begin{equation}
 d_M^{(X)} \leq \frac{P_X}{P_{\min}} -1.    \label{eq-A4}
\end{equation}

Let real numbers $t>P_{\max}+2$ and $s>1$ be parameters which will be optimized later. For given $(t,s)$, 
we select nodes $X$ and $Y$ such that $X$ is  the nearest  ancestor of $M$ under the condition $P_X>s$
and $Y$ is the nearest ancestor of $X$ under the condition $P_Y >t$. Then, in the same way as (\ref{yhani}),
we can show that
\begin{eqnarray}
 &s  < P_X   \leq 2s+P_{\max} +2, & \label{eq-A5}\\
 &t  < P_Y  \leq 2t +P_{\max} + 2. & \label{eq-A6}
 \end{eqnarray}

By combining (\ref{eq-A1-2})--(\ref{eq-A6}), we obtain the following bound of $d_M$.
\begin{eqnarray}
d_M & \leq & \log P_{\cal G} + \log \left(1 - \frac{P_{\max}+2}{P_{\cal G}} \right) 
     - \log(P_Y - P_{\max}-2) \nonumber \\
 & & \hspace{0.5cm}     + K_1[ - \log (P_{X}-1) + \log (P_Y -1) ] + K_2
 +\frac{P_X}{P_{\min}} -1   \nonumber \\
  & < & \log P_{\cal G} + \log \left(1 - \frac{P_{\max}+2}{P_{\cal G}} \right) 
     - \log(t - P_{\max}-2) \nonumber \\
 & & \hspace{0.5cm}     + K_1[ - \log (s-1) + \log (2t+P_{\max} +1) ] + K_2
 +\frac{2s+P_{\max}+2}{P_{\min}} -1  \label{eq-A7}
 \end{eqnarray}
Letting 
\begin{eqnarray}
   g(t) &=&  - \log(t - P_{\max}-2)
   + K_1 \log (2t+P_{\max} +1), \label{eq-A8}\\
   h(s) &=&  -K_1\log (s-1)+\frac{2s}{P_{\min}}, \label{eq-A9}
\end{eqnarray}
we can easily show that $g(t)$ and $h(s)$ are minimized
at $t=\tilde{t}_m$ and $s=\tilde{s}_m$, respectively, which are given by
\begin{eqnarray}
  \tilde{t}_m & =& \frac{(2+\log\alpha)P_{\max}+4+\log\alpha}{2(1-\log\alpha)}\nonumber\\
       & \approx& t_mP_{\max} + 7.676, \label{eq-A10}\\
   \tilde{s}_m& =& \frac{\log e}{2\log \alpha}P_{\min}+1 \nonumber\\
   &\approx& 1.040 P_{\min}+1, \label{eq-A11}
\end{eqnarray}
where $t_m\approx 4.405$ is defined in (\ref{argt}).
By substituting $t=\tilde{t}_m$ and $s=\tilde{s}_m$ into (\ref{eq-A7}), 
we can obtain after some calculations that 
\begin{eqnarray}
 d_M & < &   \log P_{\cal G} +(K_1-1)\log(3P_{\max}+5) - K_1 \log P_{\min} \nonumber \\
 & &  \hspace{2cm}+\frac{P_{\max}+4}{P_{\min}} + \log \left(1 - \frac{P_{\max}+2}{P_{\cal G}} \right) + K_4,\label{eq-A12}
\end{eqnarray}
where $K_4$ is defined in (\ref{eq-th5-1}). 

Since the average normalized withdrawal cost $l$ is the average of $d_M$, $l$ is bounded as follows.
\begin{eqnarray}
 l &=& \sum_{M\in{\cal G}} \frac{P_M}{P_{\cal G}} d_M \nonumber \\
   & < & \log P_{\cal G} +(K_1-1)\log(3P_{\max}+5) - K_1 \log P_{\min} 
   +\frac{P_{\max}+4}{P_{\min}} + \log \left(1 - \frac{P_{\max}+2}{P_{\cal G}} \right) + K_4\nonumber\\
 &\leq & H({\cal P}) + \log P_{\max} +(K_1-1)\log(3P_{\max}+5) - K_1 \log P_{\min} \nonumber \\
 & & \hspace{4cm} +\frac{P_{\max}+4}{P_{\min}} + \log \left(1 - \frac{P_{\max}+2}{P_{\cal G}} \right) + K_4, \label{eqA-13}
\end{eqnarray}
where the last inequality holds because  we have from (\ref{entropy}) that
\begin{eqnarray}
  \log P_{\cal G}  &=& H({\cal P}) + \sum_{M\in{\cal G}} \frac{P_M}{P_{\cal G}}\log P_M \nonumber\\
   & \leq &  H({\cal P}) +\sum_{M\in{\cal G}} \frac{P_M}{P_{\cal G}}\log P_{\max} \nonumber\\
   & = &  H({\cal P}) +\log P_{\max}.
 \end{eqnarray}


\end{document}